\documentclass[journal=ancac3,manuscript=article]{achemso}

\usepackage[version=3]{mhchem} 

\author{Etienne Gaufr\`es}
\altaffiliation{Current address: Martel group, Univ.
Montr\'eal, Montr\'eal, Canada}
\author{Nicolas Izard}
\email{nicolas.izard@u-psud.fr}
\author{Adrien Noury}
\author{Xavier Le~Roux}
\author{Gilles Rasigade}
\author{Alexandre Beck}
\author{Laurent Vivien}
\email{laurent.vivien@u-psud.fr}
\affiliation{Institut d'Electronique Fondamentale, CNRS-UMR 8622, Univ. Paris Sud,
Orsay, France}

\title{Light Emission in Silicon from Carbon Nanotubes}

\begin{document}
\begin{abstract}
The use of optics in microelectronic circuits to overcome the limitation
of metallic interconnects is more and more considered as a viable
solution. Among future silicon compatible materials, carbon nanotubes
are promising candidates thanks to their ability to emit, modulate and
detect light in the wavelength range of silicon transparency. We report
the first integration of carbon nanotubes with silicon waveguides,
successfully coupling their emission and absorption properties. A
complete study of this coupling between carbon nanotubes and silicon
waveguides was carried out, which led to the demonstration of the
temperature-independent emission from carbon nanotubes in silicon at a
wavelength of 1.3~$\mu$m. This represents the first milestone in the
development of photonics based on carbon nanotubes on silicon.

\end{abstract}

Enhancing microprocessor performances is becoming increasingly more
complex. This complexity stems from the rising transistor count and the
transistor's shrinking size in the quest to follow Moore's
Law\cite{spectrum2011}. As a consequence, power consumption in
microprocessor increases, and the on-chip communication between
differents components becomes more and more difficult, and is even now a
key-point for future multi-core generations\cite{Meindl2003}. To
overcome these problems, one of the most promising solutions is the use
of optical interconnects, which combine high data rate transmission, low
power consumption, synchronization and crosstalk\cite{IEEE-Assefa,
Zalevsky2007}. In recent years, silicon photonics was extensively
studied for the realisation of high-speed optical links\cite{Lipson2005,
Barwicz2007, Jalali2008}. Nowadays, compact photonic structures are
achieved due to the strong refractive index contrast between silicon and
silica and low loss optical propagation in the wavelengths range from
1.25~$\mu$m to 1.65~$\mu$m. In particular, various passive devices for
wavelength multiplexing or light distribution (90$^{\circ}$ bends\ldots)
are easily feasible with silicon photonics
technology\cite{Liu2010,ol-Lardenois}.

Nevertheless, silicon is an indirect bandgap material, and its
optoelectronic properties are insufficient to generate light and not
sensitive enough to detect the flux of photons transmitted in an optical
link. Other materials, such as III-V semiconductors, are good
alternatives for light emission\cite{Fang2007, Fedeli2011}, while germanium could
realize high speed photodetectors\cite{Jurgen2010, Vivien2009}. Silicon is still
used for efficient and high speed optical modulators\cite{Reed2010}. The
integration of all these materials on silicon is technically possible,
but as different and sometimes non-compatible process are used, the
resulting scheme is not cost-effective, and consequently reduce the use
of silicon photonics for a broad application domain. A monolithic
integration of laser source, optical modulator, and photodetector with a
common material would be much more favorable for emergence of photonics.

We envision the use of carbon nanotubes for all active optoelectronic
devices in silicon in order to avoid these non compatible processes.
Carbon nanotubes (SWNT) are a very versatile material, presenting at the
same time very good electronic properties\cite{Bachtold2001}, and also
optical properties\cite{Oconnell2002, Bachilo2002, Voisin2011}.  They
display strong photo- and electro-luminescence, in the 1-2~$\mu$m
wavelength range\cite{Gaufres2009, Misewich2003, Marty2006, Adam2008},
and the emission could be tuned by selecting a precise nanotube diameter
and chirality\cite{Weisman2003}. The possibility to use electrical
pumping for luminescence generation is extremely interesting for the
realization of electrically pumped optical
laser\cite{Mueller2009,Gaufres2010}.  Recent works revealed that SWNT
displays electroabsorption properties, which could be used to achieve
optical modulation\cite{Kishida2008, Izard2011}. Finally, nanotubes
present various absorption bands in the 1-2~$\mu$m range, allowing
realization of photodetectors\cite{Freitag2003, StAntoine2011}.
Therefore, carbon nanotubes are very good candidates to solve
integration issues in silicon photonics, and make cost-effective and
reliable photonics. Moreover, a side advantage of the use of nanotubes
for photonics is that the current research on the use of nanotube for
nanoelectronics\cite{ITRS} will facilitate the integration between
photonics and electronics.

In this paper, we propose a way to integrate nanotubes with silicon
photonics technology, which is a key-point for future realisation of
carbon nanotube based photonic devices. This work looks into the
difficulties of coupling the optical properties of a 1D nanomaterial
(SWNT) to a bulk 3D material like silicon. In this context, we report
the first integration of nanotubes' absorption and emission properties
in silicon waveguides at telecommunication wavelengths around 1.3~$\mu$m.

\section{Results and discussions}

\subsection{Integration scheme design}

The integration scheme considered is based on the insertion of an
interacting zone in an input/output silicon waveguide. It is constituted
of a sub-micron silicon waveguide embedded in a carbon nanotubes
composite, as schematically presented in Fig. 1. The input and output
waveguides have both a width $W_g$ of 450~nm. For such geometry, the
optical mode is strongly confined into the waveguide (Fig. 1A) due to
the high refractive index contrast between silicon (optical index 3.45)
and its top and bottom cladding layers where the index is near 1.45.
This confinement prevents strong interactions with the surrounding
media, allowing high transmission of optical signal. On the other hand,
the waveguide in the interaction region with carbon nanotubes has a
width $W_i$ below the critical minimum confinement width $W_c$, which is
the limit under a significant fraction of the optical mode starts to
leak outside the silicon waveguide. At 1.3~$\mu$m, $W_c$ is around
400~nm. A judicious shrinkage of the waveguide width $W_i$ allows both
deconfining a controlled fraction of the optical mode and preserving the
propagation of the optical wave along the waveguide. Fig. 1B illustrates
the mode confinement calculated for a $W_i$ of 275~nm and shows the
optical mode spreading outside the waveguide silicon core. As a
significant fraction of the energy is propagating outside the waveguide,
the guided mode will have a strong interaction with the active SWNT
based polymer top cladding layer. To minimize optical losses in the
transition between the input/output waveguide ($W_g$ = 450~nm) and the
interaction region waveguide ($W_i$ = 275~nm), adiabatic tapers are
required (typically 300~$\mu$m long).

\subsection{Technology}

\begin{figure}
  \includegraphics[width=14cm]{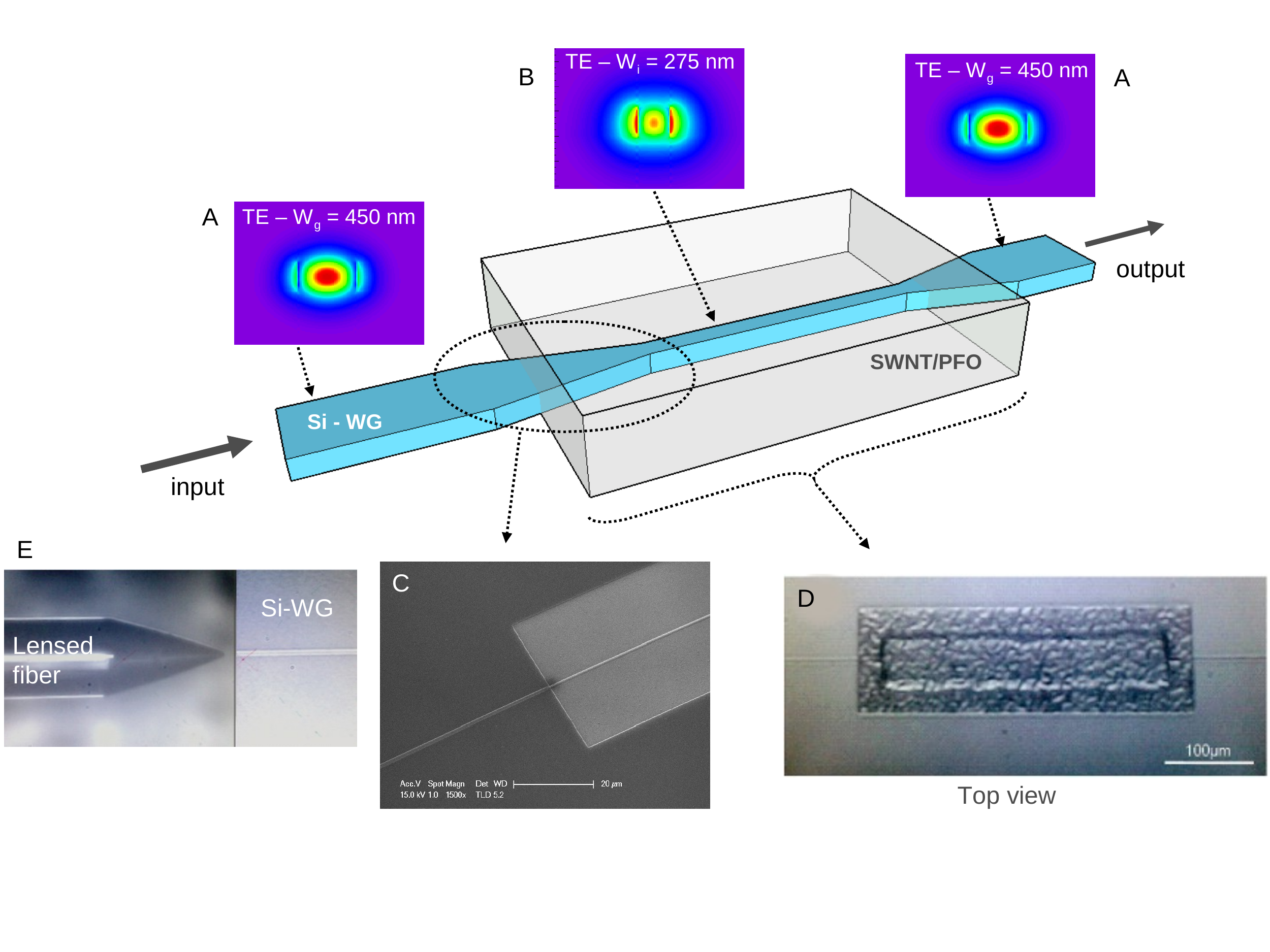}
  \caption{Integration scheme of carbon nanotubes thin layer composite
with silicon waveguide. (a) Input and output single-mode silicon
waveguides with a height of 220~nm and a width $W_g$ of 450~nm. The
optical mode for TE polarization is strongly confined into the silicon
waveguide. (b) Interaction region with carbon nanotubes and the silicon
waveguide, with a height of 220~nm and a width $W_i$ of 275~nm. The
guided mode is deconfined, and a significant fraction of the energy 
propagate outside the waveguide. (c) SEM view of the adiabatic tapper
which adapt the mode from the input waveguide ($W_g$) to the interaction
zone waveguide ($W_i$). (d) Optical microscope top view of the final
device, with the silica recess and SWNT/PFO thin layer on top of it. The
apparent roughness on top of the SWNT/PFO thin film is due to the low
temperature \ce{SiO2} hard mask used in the last technological step. (e)
Optical microscope view of the lensed fiber used to inject the input
light into the cleaved waveguide facet.}
  \label{fig1}
\end{figure}

The substrate is a silicon-on-insulator (SOI) with a 220~nm thick
silicon layer on top of a 1~$\mu$m thick buried silica layer.  Although
details of the technological steps are given in supporting information,
briefly, silicon waveguides were made using e-beam lithography
patterning followed by reactive ion plasma etching (RIE). A 500~nm thick
silicon dioxide (\ce{SiO2}) protection layer was deposited onto the
silicon wafer using a plasma-enhanced chemical vapor deposition (PECVD)
technique. A silica recess of length $L$ corresponding to the
interaction region was etched down to the buried silicon oxide layer in
order to expose the vicinity of the waveguide allowing interactions with
the surrounding medium, and in particular SWNT. (\ref{fig1}C).

The single-wall carbon nanotube (SWNT) film was prepared as follow:
as-prepared HiPCO SWNT powder (Unydim Inc.) were mixed with
poly-9,9-di-n-octyl-fluorenyl-2,7-diyl (PFO) in toluene at a ratio of
SWNT (5~mg): PFO (5~mg): toluene (30~ml). This mixture was homogenized
by sonication (for 1~h using a water-bath sonicator and 15~min using a
tip sonicator) and was centrifuged for 5-60~min using a desktop
centrifuge (angle rotor type, 10,000~g). Solution was then drop casted
directly into the previously defined silica recess to form a 1~$\mu$m
thick layer. Samples were further annealed at 180$^\circ$C for 15~min
to improve optical quality of the SWNT/PFO
film\cite{Gaufres2009,ox-Gaufres}. This layer was then delimited around
the interaction region by another \ce{O2}-based RIE to remove the
SWNT/PFO film on the undesired areas (\ref{fig1}D).

\begin{figure}
  \includegraphics[width=14cm]{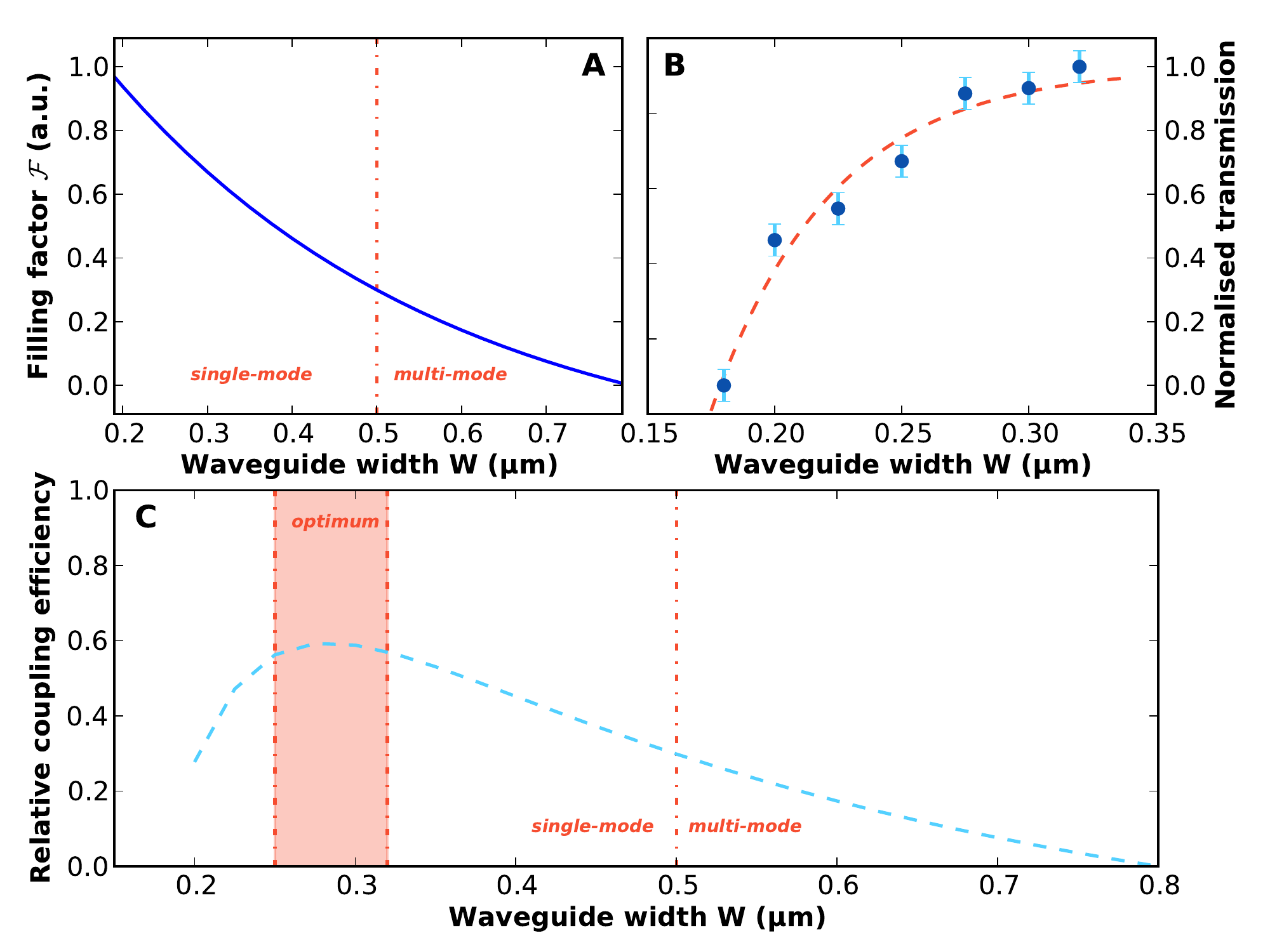}
  \caption{Determination of the waveguide width $W$ to optimize the
interaction between the optical mode in silicon waveguide and carbon
nanotubes. (a) Simulation of the evolution of the optical mode fraction
confined in SWNT/PFO layer (filling factor $\mathcal{F}$) as a function
of the silicon waveguide width $W$. (b) Experimental results of the
normalized transmission as a function of the silicon waveguide width
$W$. (c) Optimal conditions which fullfill low loss and strong
interaction of the optial modes with SWNT. This results from the
convolution of (a) and (b). The optimum waveguide with range is between
250~nm to 330~nm.}
  \label{fig2}
\end{figure}

The evolution of the optical mode fraction confined in the SWNT/PFO
layer (filling factor $\mathcal{F}$) was determined as a function of the
waveguide width $W$ by mode-solving numerical simulations using an
home-made simulation code, and are displayed in \ref{fig2}A. The
multimode/singlemode limit is found to be around 500~nm. As $W$ narrows
down, the filling factor (\textit{i.e.} the amount of energy in the SWNT/PFO
layer) increases, thus strengthening optical mode interaction with
carbon nanotubes. Ideally, the waveguide width $W$ should be reduced
down to 150~nm and even less to optimize light interaction with SWNT.
However, silicon waveguides are on top of a 1~$\mu$m thick \ce{SiO2}
layer. If the optical mode transmitted into the waveguide is too
deconfined, it will start to leak throught the \ce{SiO2} layer towards
the silicon substrate, resulting in huge losses. \ref{fig2}B displays
experimental transmission results performed on optical waveguide of the
same length with several width $W$. As $W$ is reduced, the optical
transmission throught the waveguide exponentially decreases.

There is an optimum for a waveguide width to fullfill both high
transmission and strong interactions with SWNT/PFO layer. This optimum
could be determined thanks to the convolution of the experimental
transmission and the simulated filling factor. The obtained result is
displayed in \ref{fig2}C. The optimum waveguide width $W$ was found
to be between 250 and 330~nm. Consequently, a waveguide width $W_i$ of
275~nm was used for the interaction region.

\subsection{Absorption coupling}

The integration of SWNT in silicon and the coupling of their optical
properties was first demonstrated by measuring the absorption of the
upper SWNT/PFO layer across the waveguide. A laser beam from a tunable
fibered laser source with an emission wavelength centered at 1.3~$\mu$m
was used at TE polarization (\textit{i.e.} electric field parallel to the
substrate). Input light was injected into the cleaved waveguide facet
using a lensed fiber (\ref{fig1}E). A liquid nitrogen cooled InGaAs
monocanal detector recorded the output beam through a monochromator.

\begin{figure}
  \includegraphics[width=14cm]{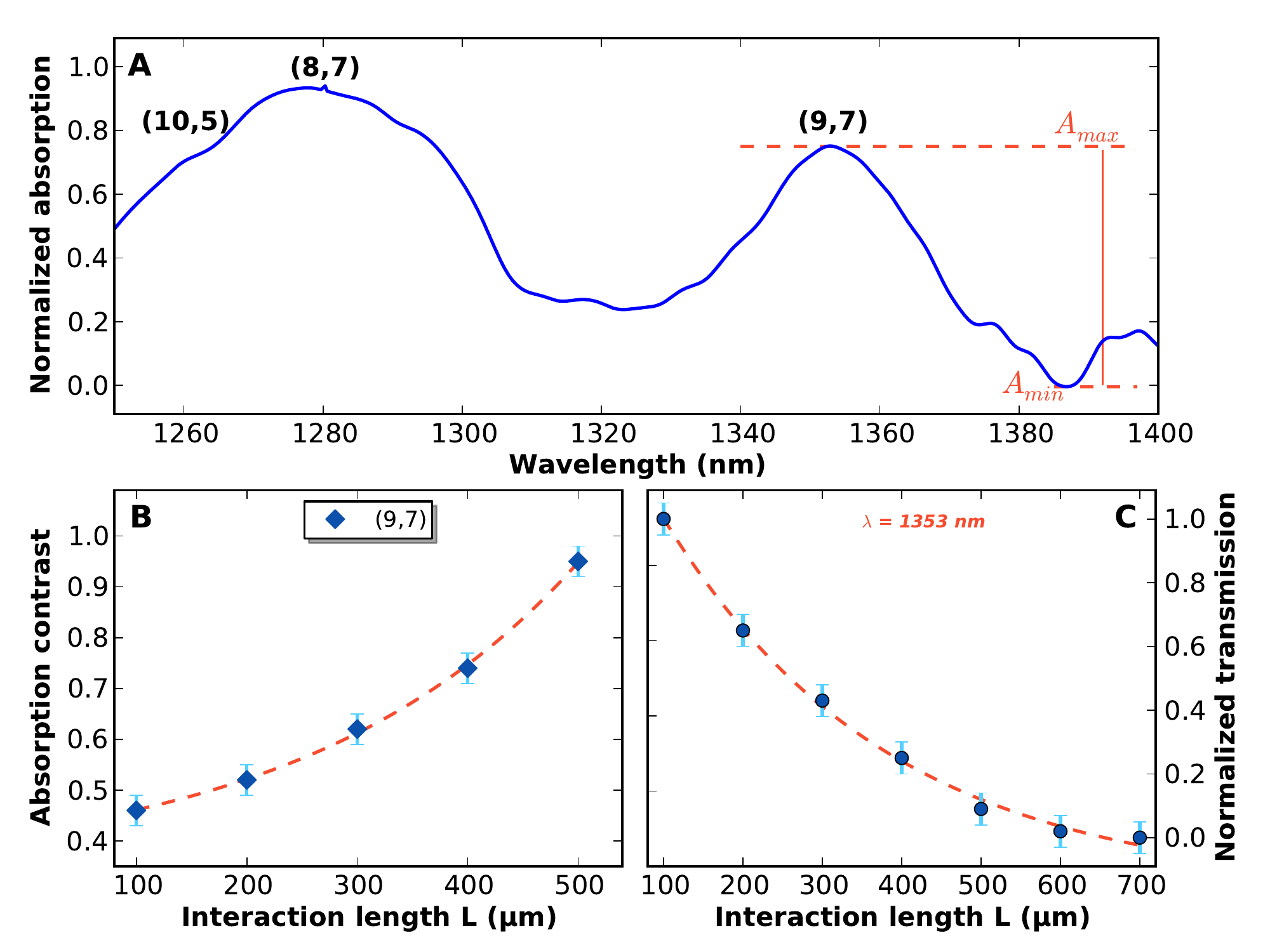}
  \caption{Carbon nanotube absorption measurements in silicon
waveguide. Evolution with the interaction length $L$. (a) Normalized
absorption as a function of wavelength from 1250~nm to 1400~nm.
($10$,$5$), ($8$,$7$) and ($9$,$7$) species are clearly observed. (b)
Absorption contract $C$ of the ($9$,$7$) nanotube, defined by
$\frac{A_{max}-A_{min}}{A_{max}+A_{min}}$ as a function of the
interaction length $L$. (c) Normalized transmission of the ($9$,$7$)
nanotube at $\lambda = 1353$~nm as a function of the interaction length
$L$.}
  \label{fig3}
\end{figure}

\ref{fig3}A reports the absorption spectrum of the carbon nanotubes
through the silicon waveguide. In this case, the waveguide width $W_i$
and length $L$ in the interaction region were 275~nm and 400~$\mu$m
respectively. Several absorption peaks could be observed, each one
corresponding to the specific absorption of one kind of nanotubes with
specific ($n$,$m$) index. In the considered wavelength range
(1.25-1.4~$\mu$m), three differents SWNT are clearly identified,
corresponding to ($10$,$5$), ($8$,$7$) and ($9$,$7$) indexes. 

The influence of the interaction length $L$ was studied from the
absorption contrast $C$ of the ($9$,$7$) SWNT defined as:

\begin{equation}
	C = \frac{A_{max}-A_{min}}{A_{max}+A_{min}}
\end{equation}

where, $A_{max}$ and $A_{min}$ are respectively the maximum and minimum
absorption of the ($9$,$7$) absorption peak, as defined on \ref{fig3}A.

Both the absorption contrast and the output intensity level were
determined for several waveguide interaction lengths $L$, increasing the
ranging from 100 to 500~$\mu$m. Results are displayed in \ref{fig3}B and
\ref{fig3}C, respectively. For the same waveguide width $W_i$ of 275~nm,
the absorption contrast $C$ increase when $L$ increase. On the same
time, we found out that the waveguide transmission at $\lambda =
$~1353~nm (corresponding to the absorption peak of the ($9$,$7$)
nanotube) is decreasing when $L$ increase. Both results follow a
Beer-Lambert type exponential law, indicating that the nature of the
interaction between the optical mode propagated throught the waveguide
and carbon nanotubes is, indeed, optical absorption and not diffusive in
nature. This result is the first signature of an effective coupling
between SWNT absorption and the optical mode transmitted by the silicon
waveguide.

Considering integration of carbon nanotubes with future photonic
devices, there is a trade-off between the interaction with nanotubes
(absorption contrast) and the waveguides transmission level. Several figures
of merit could be proposed depending on the relative weight between
absorption contrast and the global transmission of the device, however,
an optimum interaction length $L_i$ would be in the range 400 to
500~$\mu$m, leading to very compact and low loss SWNT based photonic devices.

\subsection{Light emission}

\begin{figure}
  \includegraphics[width=14cm]{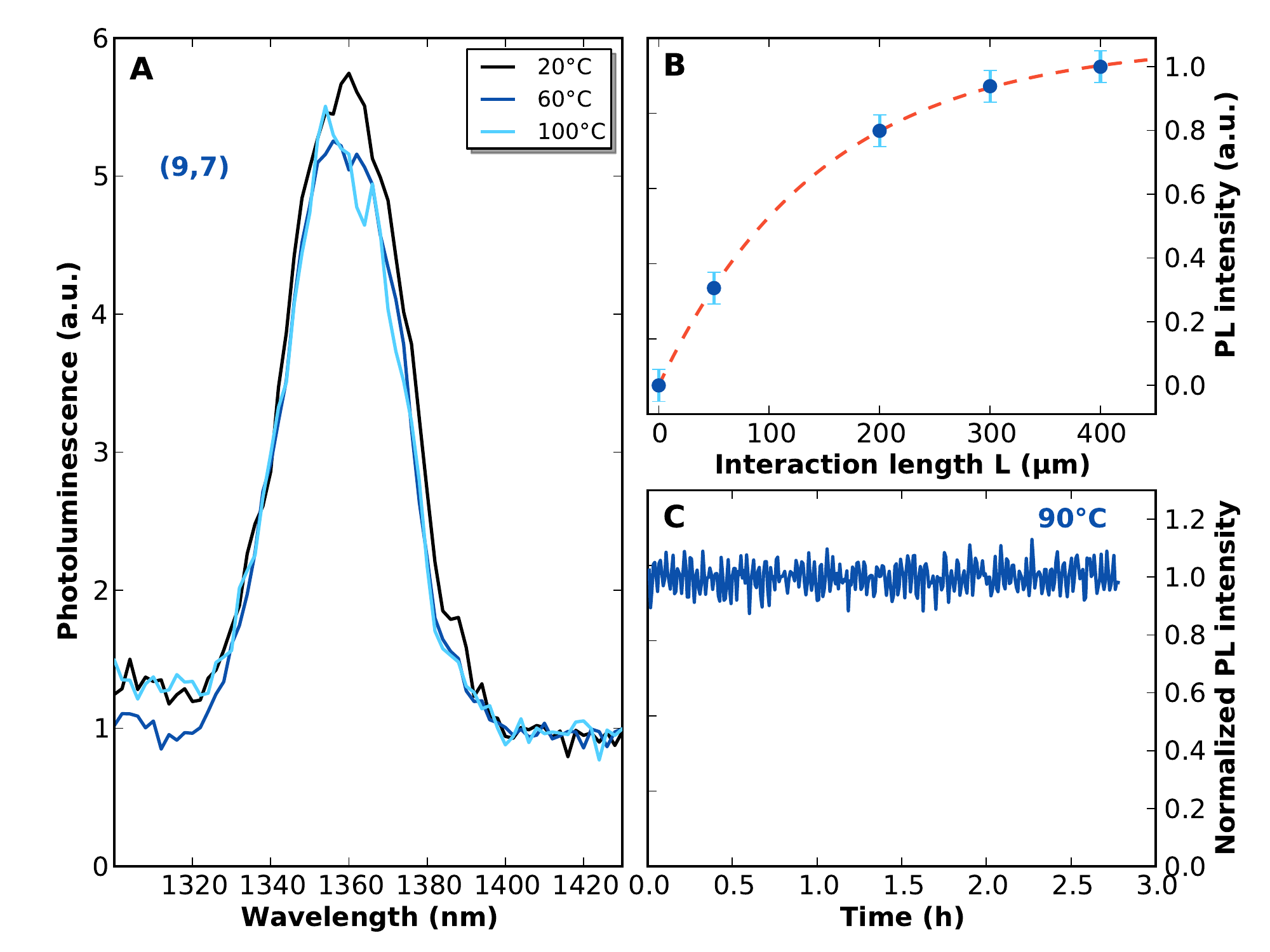}
  \caption{Coupling of carbon nanotubes PL in silicon waveguide. (a)
Photoluminescence generated from the ($9$,$7$) nanotube under excitation
by a 800~nm pump laser, coupled in silicon waveguide for three
temperatures (20$^\circ$C, 60$^\circ$C and 100$^\circ$C). No wavelength
shift or signal reduction are observed. (b) Evolution of the photoluminescence
intensity as a function of the interaction waveguide length $L$. (c)
Time stability of the photoluminescence coupled in silicon waveguide at
90$^\circ$C.}
  \label{fig4}
\end{figure}

In order to investigate whether or not the proposed integration scheme
is promising for carbon nanotube based optical sources, the light
emission from SWNT into silicon waveguides was studied. A laser diode
emitting at a wavelength of 802~nm was used to optically pump SWNT, in
particular the (9,7) nanotube population. The incident laser beam
arrived from the top of the silicon die, and was focused on the silica
recess, where carbon nanotubes could interacts with the silicon
waveguide. The light outgoing from the output waveguide facet, placed
few millimeters away from the interaction region, was collected using a
20X microscope objective with 0.35 numerical aperture. The spectrum is
displayed in \ref{fig4}A.

This (9,7) SWNT presents a strong photoluminescence (PL) peak, at a
wavelength around 1.35~$\mu$m, which is identical to the one reported in
a thin PFO layer\cite{Gaufres2009}. The influence of the interaction
length $L$ on the photoluminescence was also studied for several lengths
between 50~$\mu$m and 400~$\mu$m. Results are displayed on \ref{fig4}B.
For clarity, a point with an interaction length $L = $~0~$\mu$m and PL
intensity of 0 was added to results. We noticed that the increase in PL
intensity when the interaction length $L$ increase is not linear.
Indeed, its intensity increase until it reaches a maximum at $L$ around
400~$\mu$m. This evolution could be described by a Beer-Lambert type
exponential law, meaning that the observed limitation in
photoluminescence intensity is due to the SWNT/PFO layer secondary
absorption. These results are consistent with previously obtained
results for the absorption coupling and suggest that the optimal
interaction length $L_i$ for SWNT/PFO based photonic devices, relying on
SWNT absorption or emission properties, would be around 400~$\mu$m.

In order to furthermore determine the strength of carbon nanotubes for
silicon photonic applications, we determined the SWNT emission stability
with temperature. That is a major requirement for the achievement of
most applications. Evolution of the (9,7) nanotubes emission through the
silicon waveguide is presented in \ref{fig4}A for a given interaction
length $L_i$ of 400~$\mu$m. We notice that SWNT PL peak did not shift in
wavelength and keep the same intensity level for temperature ranging
from 10$^\circ$C to 100$^\circ$C, which is typically the operating
temperature of photonic and microelectronic systems. In order to assess
thermal stability over time, PL intensity of ($9$,$7$) SWNT was recorded
for 3 hours at 90$^\circ$C, and is displayed in \ref{fig4}C. The
emission through the waveguide remained constant, and no thermal
degradation or intensity decrease was observed.

One important point in the coupling of the photoluminescence of a 1D
material such as SWNT into the optical mode of a bulk 3D material like a
silicon waveguide, is the coupling efficiency $\Psi$, which is the
number of photon effectively transmitted through the waveguide over the
total number of photon emitted by carbon nanotubes. That is:

\begin{equation}
	\Psi = \frac{N_{photons\ transmitted}}{N_{photons\ emitted}}
\end{equation}

However, this is a very difficult task to undertake in this case.
Indeed, $N_{photons\ emitted}$ is related to SWNT quantum yield
$\Phi$, where:

\begin{equation}
	\Phi = \frac{N_{photons\ emitted}}{N_{photons\ absorbed}}
\end{equation}

Unfortunately, $\Phi$ is not a well-known parameter in carbon nanotubes,
and it may vary greatly, in the range $10^{-4}$ to $10^{-1}$, depending
of SWNT diameter, surrounding or length\cite{Cognet2008, Hertel2010}. In
particular, $\Phi$ is not yet determined for PFO embedded SWNT.

In any case, it it possible to determine the \emph{external effective coupling
factor} $Q_{eff}$, where:

\begin{equation}
	Q_{eff} = \Phi \cdot \Psi = \frac{N_{photons\ transmitted}}{N_{photons\ absorbed}}
\end{equation}

Knowing the input and output power, the SWNT/PFO thin layer absorbance
and the collecting setup transmission, $Q_{eff}$ was estimated to be
$2 \cdot 10^{-5}$.

If we estimate that carbon nanotubes quantum yield is low (\textit{e.g.} $\Phi =
10^{-4}$), this means that coupling efficiency $\Psi$ is of the order of
$10^{-1}$. In the future, the use of photonic cristals or slot
waveguides might increase the coupling between carbon nanotubes and
silicon waveguides.

In light of the results obtained so far, one could state that carbon
nanotubes is a promising material for development of new compact and
temperature independent photonics devices on silicon.

\section{Conclusions}

In conclusion, the integration of carbon nanotubes properties
(absorption and photoluminescence) in silicon waveguides was studied
using an integration scheme based on evanescent waveguides. The coupling
of carbon nanotubes photoluminescence into silicon waveguides has been
demonstrated over a wide temperature range. The external effective
coupling factor $Q_{eff}$ was estimated about $2 \cdot 10^{-5}$. In
the future, improved integration schemes based on photonic cristals or
slot waveguides will be considered to further increase carbon nanotubes
coupling with photonic waveguides in order to fully exploit their
extraordinary optical properties to achieve efficient optoelectronic
devices in silicon.  

\section{Materials and Methods}
\subsection{Carbon nanotubes}

Single-wall nanotube powder was purchassed from Unydim Inc. These
nanotube were fabricated using the HiPCO process, and were not purified.
poly-9,9-di-n-octyl-fluorenyl-2,7-diyl (PFO) was purchassed from Sigma
Aldricht. SWNT (5~mg) and PFO (5~mg) were mixed in toluene (30~ml). The
mixture was homogeneized, first using a water-bath sonicator for 1h,
second, using a tip sonicator at 20~\% power for 15~min. In order to
remove a part of metallic catalyst particles from the HiPCO process and
other impurities, this mixture was then centrifugated using a desktop
centrifuge (typically 5-60~min at 10.000~g). The supernatant is then
collected and could be drop casted directly onto the photonics devices.
The thin film are annealed at 180$^{\circ}$ for 15 min to improve their
optical quality.

\subsection{Waveguide fabrication}

Silicon-on-insulator (SOI) substrate were purchassed from Soitec, with a
top silicon layer of 220~nm, and a buried \ce{SiO2} layer of 1~$\mu$m.
The technological steps for waveguide fabrication are as follow (see
also Supplementary Materials S1):
\begin{itemize}
\item Step 1: SOI substrate preparation: Cleaning with acetone under sonication,
followed by \ce{O2} plasma cleaning.

\item Step 2: Waveguides definition into the e-beam resist. Deposition
of 300~nm thick MaN negative resist. E-beam patterning at 20~keV, 30~s
resist developing in MIF~726.

\item Step 3: Pattern transfert: Etching of silicon by RIE down to 6~nm
thick silicon layer. RIE: \ce{SF6} (20~sccm) and \ce{O2} (5~sccm) at
30~W for 70~s.

\item Step 4: Waveguide protection by \ce{SiO2}: 500~nm thick \ce{SiO2}
layer is deposited on top of silicon waveguides at 300$^\circ$C.

\item Step 5: Interaction zone definition into the e-beam resist.
Deposition of 300~nm thick positive ZEP resist. E-beam patterning at
20~kEV. Developing: 30~s ZED50 / 30~s MIBK / 30~s IPA.

\item Step 6: Interaction zone opening: Wet etching with diluted HF of
silica down to silicon stopping layer. \ce{O2} plasma for resist
removal.

\item Step 7: SWNT/PFO deposition: 1~$\mu$m thick SWNT/PFO layer is
deposited on the full die followed by a thermal annealing at
180$^\circ$C.

\item Step 8: SWNT/PFO protection with \ce{SiO2}: 200~nm thick \ce{SiO2}
layer is deposited on top of the SWNT/PFO layer at low temperature
(150$^\circ$C) to prevent layer damage.

\item Step 9: Definition of protected zone into the e-beam resist. MaN
resist and e-beam patterning, developing.

\item Step 10: \ce{SiO2} etching using RIE : \ce{CHF3} (50~sccm) and
\ce{O2} (3~sccm) at 325 W (12~mTorr) for 190~s.

\item Step 11: Removal of SWNT/PFO outside defined zone by RIE etching.
50~\% Ar and 50\% \ce{O2} at 300 W for 15~min.

\end{itemize}

\subsection{Simulations}

Optical simulations were performed using an home-made 2D mode solver
based on the full-vectorial finite-difference
method\cite{IEEE2008} and focused on anisotropic dielectric waveguides.

The filling factor $\mathcal{F}$, measure of the fraction of the mode
power flux in the SWNT/PFO layer, is defined by :

\[
\mathcal{F} = \frac{\int^R P(s)\,\mathrm{d}s}{\int^\infty P(s)\,\mathrm{d}s}
\]

where $R$ is the SWNT/PFO layer and $P(s)$ is the mode power flux.

\subsection{Determining $Q_{eff}$}

Incoming power density on the silica recess $J_i$ was estimated to be
$2.7 \cdot 10^6$~W$\cdot$m$^2$. A simple assumption is made considering the area around
the waveguide whether a nanotube could interact or not. As the profile
of the evanescent field is gaussian, it is considered that the whole
energy of the optical mode is confined within $5\sigma$ of the gaussian.
That is, a nanotube inside $5\sigma$ \emph{may} interact with the
waveguide, while a nanotube outside $5\sigma$ \emph{will not} interact
with the waveguide. Consequently, only the SWNT/PFO thin layer surface
inside this $5\sigma$ limit will be considered.
For a waveguide width $W_i$ of 275~nm, the $5\sigma$ limit is: 360~nm
(cf. Supplementary Materials S2). That is, the surface area of the SWNT/PFO layer is
$S_{NT}$ = $1.1 \cdot 10^{-11}$~m$^2$.
The incident pump power on carbon nanotubes which could couple into the
waveguide is $P_i = J_i \cdot S_{NT} = 2.91 \cdot 10^{-5}$~W.
Absorption $A$ of a 1~$\mu$m thick SWNT/PFO layer at 805~nm is $6.6
\cdot 10^{-2}$ (cf. Supplementary Materials S3). Absorbed power in carbon
nanotubes $P_a = A \cdot P_i = 1.92 \cdot 10^{-6}$~W.

On the other hand, power of the collected photoluminescence from the
waveguide was $P_c = 1.27\cdot10^{-11}$~W. The collecting setup losses
were estimated to be 2~dB. So the transmitted power was $P_t = P_c \cdot
10^{0.2} = 2.01\cdot10^{-11}$~W.

The external effective coupling factor $Q_{eff}$ could be determined by:


\[
Q_{eff} = \frac{N_{photons\ transmitted}}{N_{photons\ absorbed}} = 
\frac{P_t / E_t}{P_a / E_a} =
\frac{P_t}{P_a} \cdot \frac{\nu_a}{\nu_t} = \frac{P_t}{P_a} \cdot
\frac{\lambda_t}{\lambda_a}
\]

where $E$ is the energy, $\nu$ the frequency and $\lambda$ the
wavelength of the transmitted and absorbed photons.

Finally, $Q_{eff} = 1.8\cdot10^{-5} \simeq 2\cdot10^{-5}$.

\acknowledgement
Authors would like to thank N. Tang from Univ. Montr\'eal for her
proofreading of the manuscript, and D. Marris-Morini and E. Cassan
from Univ. Paris Sud for fruitfull discussions.

\suppinfo
Detailled technological process of the waveguide fabrication (S1),
evanescent mode profile of a 275~nm with waveguide (S2), and absorption
spectra of a 1~$\mu$m thick SWNT/PFO thin layer on silica (S3). This material
is available free of charge \textit{via} the internet at http://pubs.acs.org

\bibliography{integrated}

\end{document}